\documentclass[num-refs,manuscript=article,layout=twocolumn]{1-article}

\usepackage{hyperref}
\usepackage{doi}
\usepackage{ragged2e} 
\usepackage[version=4]{mhchem} 
\usepackage[english]{babel}
\usepackage{lipsum}
\usepackage[version=4]{mhchem}
\usepackage{parskip}
\usepackage{amsmath}
\usepackage[markup=underlined]{changes}
\papertype{Research Article}
\usepackage{placeins}

\begin{document}

\title{Smart sensing of the multifunctional properties of magnetron sputtered \ce{MoS2} across the amorphous-crystalline transition.}

\author[1]{Jose L. Ocana-Pujol}
\author[1]{Rebecca A. Gallivan}
\author[1]{Ramon Camilo Dominguez Ordonez}
\author[1]{Nikolaus Porenta}
\author[2]{Arnold M\"uller}
\author[2]{Christof Vockenhuber}
\author[1]{Ralph Spolenak}
\author[1]{Henning Galinski}

\affil[1]{
Laboratory for Nanometallurgy, ETH Zurich, Z\"urich, Switzerland}
\affil[2]{Laboratory of Ion Beam Physics, ETH Zurich, Z\"urich, Switzerland}

\corraddress{Department of Materials, Laboratory for Nanometalurgy, Zurich, 8093, Switzerland}
\corremail{joseo@ethz.ch and henningg@ethz.ch}
\fundinginfo{ETH Research. Grant Number: ETH-47 18-1}


\maketitle

\begin{abstract}
Molybdenum disulfide, \ce{MoS2}, is a next-generation semiconductor and is frequently integrated into emergent optoelectronic technologies based on two-dimensional materials. Here, we present a method that provides direct optical feedback on the thickness and crystallinity of sputter-deposited \ce{MoS2} down to the few-layer regime. This smart sensing enables tracking the material's functional properties, such as excitonic response, sheet resistance, and hardness across the amorphous-crystalline transition. To illustrate the potential of such feedback-controlled fabrication, we realized \ce{MoS2}-based hyperbolic metamaterials (HMM) with controllable optical topological transitions and hardness.
\end{abstract}

\newpage
\section{Introduction}
Semiconductors are essential building blocks of the electronic devices that critically shape our daily lives~\cite{history_semiconductor}. However, the drive for device miniaturization faces challenges as we approach the fundamental physical limits of today's semiconductor technology~\cite{wang_road_2022}. In this context, molybdenum disulfide (\ce{MoS2}), a transition metal chalcogenide (TMC), has emerged as a potential next-generation semiconductor. Unlike other TMCs, \ce{MoS2} is composed of relatively abundant and relatively innoxious elements~\cite{gupta_comprehensive_2020} and has a long history of being used in industry due to its catalytic~\cite{wilcoxon_optical_1994,kadiev_mechanism_2021} and mechanical properties~\cite{spalvins_bearing_1975,rapoport_friction_2008}. These properties stem from the material's peculiar layered crystal structure, in which each layer consists of a molybdenum sheet sandwiched between two sheets of sulfur atoms. While in the plane these layers form a covalently bonded hexagonal lattice, they are only held together by weak van der Waals forces on the interlayer level~\cite{yazyev_mos2_2015}.
\par
As a consequence of weak interlayer bonding, single- to few-layer~\ce{MoS2} is readily produced by top-down approaches such as mechanical or chemical exfoliation. Similar to graphene, the electronic properties of \ce{MoS2} change dramatically when the crystal is reduced to a single layer~\cite {mak_evolution_2010}. While bulk crystals have an indirect band gap ($1.23$~eV) requiring phonon interaction for bound electron-hole pair (exciton) formation, in the monolayer limit, \ce{MoS2} exhibits a direct bandgap ($1.89$~eV), featuring strong excitonic transitions~\cite{mak_atomically_2010}. Its tunable electronic properties make \ce{MoS2} a promising material for various classical semiconductor applications such as transistors~\cite{kim_infrared_2019}, light-emitting diodes~\cite{hwangbo_wafer-scale_2022}, solar cells~\cite{al-ghiffari_systematic_2022}, and emergent optoelectronic applications such logical gates based in spin- and valleytronics~\cite{mak_lightvalley_2018,zhang_enhancing_2019,ahn_2d_2020}. 
\par
Despite the tempting prospects of \ce{MoS2} as a next-generation semiconductor, the integration of \ce{MoS2} into standard semiconductor fabrication processes remains a significant challenge. Although developments in top-down approaches such as exfoliation have proven capable of overcoming initial drawbacks in coverage and scalability, these techniques remain mainly limited to monolayer systems~\cite{chang_fast_2020,quellmalz_large-area_2021}. Similarly, bottom-up approaches such as chemical vapor deposition (CVD) require high annealing temperatures ($T>1000$ K), limiting the choice of substrate materials on which \ce{MoS2} can be grown~\cite{timpel_2d-mos2_2021,quellmalz_large-area_2021}. All the mentioned techniques can only produce crystalline \ce{MoS2} (\ce{c-MoS2}), leaving the properties of amorphous \ce{MoS2} (\ce{a-MoS2}) not fully explored ~\cite{cao_tuning_2017,krbal_amorphous--crystal_2021}. Notably, \ce{a-MoS2} shows superior optical broadband absorbance~\cite{huang_amorphous_2019} and hydrogen evolution reaction activities compared to \ce{c-MoS2}, making it more suitable for applications such as photodetectors and hydrogen catalysis~\cite{huang_amorphous_2019,wu_HER_MoS2}. 
\par
To address these challenges, magnetron sputtering, which is a bottom-up fabrication technique well aligned with standard semiconductor fabrication processes, has been considered as an alternative route to fabricate \ce{MoS2}. Sputtering enables the production of both amorphous and crystalline \ce{MoS2} with thicknesses ranging from one monolayer to more than one micrometer~\cite{tao_growth_2015,babuska_quality_2022} on all vacuum-compatible substrates. Still, common drawbacks of sputtering are the high kinetic energy upon arrival, which is typical of the sputtering process, leading to a limited degree of epitaxy and an increase in defects in the material. Hence, to establish a fabrication procedure for sputtered \ce{MoS2} with a high uniformity and controllable number of layers, it is imperative to develop tools that enable effective monitoring of the thickness, crystal structure, and uniformity post-deposition.
\par
In this work, we design a low-footprint thermochromic sensor to assist in the controlled deposition of amorphous and crystalline \ce{MoS2} by magnetron sputtering. The sensor relies on an asymmetric Fabry-Pérot-type resonator~\cite{kats_nanometre_2013} that utilizes the optical losses in \ce{MoS2}~\cite{islam_-plane_2021} to optically detect changes in thickness and crystal structure down to a monolayer. This in-place monitoring enables us to study the chemical and functional properties of \ce{MoS2}, such as exciton formation, hardness, and sheet resistance, across the amorphous-crystalline phase transition. We highlight the advantage of direct optical feedback within the fabrication process by realizing a \ce{MoS2}-based hyperbolic metamaterial and analyzing its functional properties. 
\section{Results and Discussion}
\subsection{Monitoring the bottom-up fabrication of \ce{MoS2}}
\begin{figure*}[ht]
\begin{center}
  \includegraphics[width=0.9\linewidth]{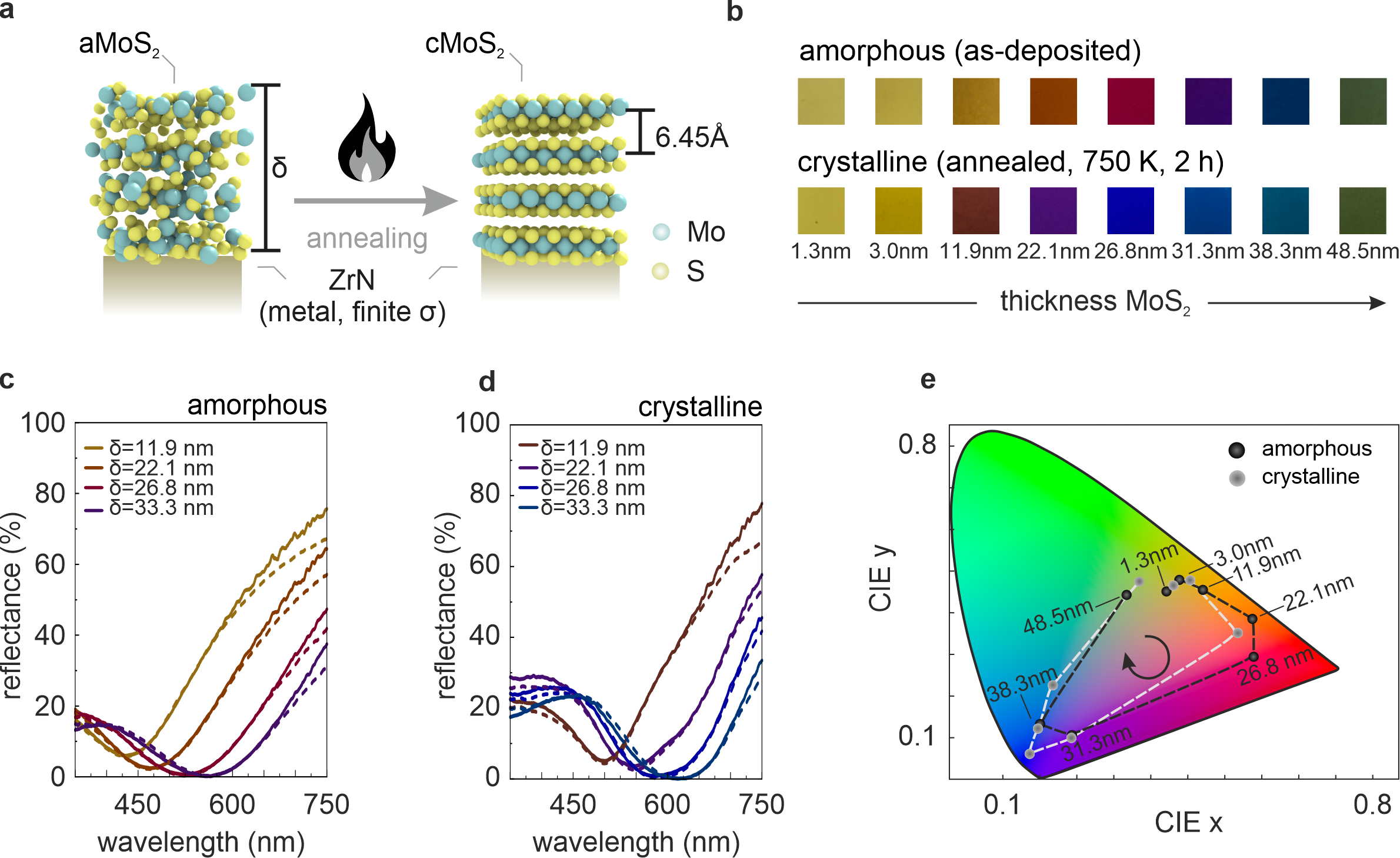}
  \caption{(a): Schematic illustration of the thermochromic sensor in the amorphous and crystalline states. \ce{ZrN} acts as an inert metallic backreflector and \ce{MoS2} is the cavity medium. (b): Photographs of \ce{MoS2} on top of \ce{ZrN}-coated Si wafers ranging from 2-layers of \ce{MoS2} to bulk \ce{MoS2} before (top) and after annealing (bottom). The indicated thickness $\delta$ was measured by RBS. (c) and (d): Normal incidence reflectance spectra of \ce{MoS2} on top of \ce{ZrN} before and after annealing. (e): CIE 1931 color map showing the colors achieved in the samples in (b). The dark symbols indicate the colors of the amorphous samples, while the grey symbols indicate the colors of the crystalline samples.}
  \label{fig:MoS2_1}
\end{center}
\end{figure*}\FloatBarrier
To identify the operational regime of the thermochromic sensor, \ce{a-MoS2} of varying thickness $\delta$ was deposited on zirconium nitride (\ce{ZrN}) by RF magnetron sputtering. In a second step, crystallization is achieved by annealing at $750$~K for two hours in ultrahigh vacuum (UHV), as illustrated in Figure~\ref{fig:MoS2_1}~a. Here, \ce{ZrN} acts as a metallic back-reflector of the asymmetric cavity. \ce{ZrN} was specifically selected as it guarantees high chemical stability in a sulfur-rich environment. We refer to the supporting information (Figure~S1) for a more detailed discussion on the selection of a metallic backreflector.
\\~\\
Figure~\ref{fig:MoS2_1}~b shows optical micrographs of amorphous and crystalline \ce{MoS2} directly deposited on \ce{ZrN}-coated Si wafers. In both states, we observe the formation of vivid structural colors spanning from yellow to green, depending on the thickness. This change in coloration is also seen in the reflectance spectra obtained by near-normal incidence reflectometry shown in Figure~\ref{fig:MoS2_1}~c and d. The reflectance minimum observed in the spectra linearly red-shifts on average by $6.4 \pm 0.5$~nm (Figure~S2) due to an increase in thickness $\delta$. The experimentally measured spectra are in good agreement with calculated spectra based on Transfer Matrix calculations~\cite{linton2009wave,byrnes_multilayer_2020} (Figure~\ref{fig:MoS2_1}~c and d). Refractive indexes were determined using ellipsometry (Figure~S3, Supporting Information). 
\\~\\
Upon crystallization, both the optical properties and structural coloration of the sputtered \ce{MoS2} films change(Figure~\ref{fig:MoS2_1}~d). The observed strong thermochromic response is due to the high absorption in the asymmetric optical cavity. The spectral position of the absorbing state (i.e., minimal reflectance) red-shifts on average by $45 \pm 25$~nm for a given thickness due to an increase in refractive index upon crystallization (Figure~S3), as the resonance condition scales with $\lambda_{\text{min}}\propto n_{\ce{MoS2}}\cdot \delta$.
\\~\\
While the thermochromic response can be identified by eye in the $11.9-38.3$~nm thickness range, all configurations exhibit a measurable change in color as shown in the CIE color plot (Figure~\ref{fig:MoS2_1}~e). For both parameters, namely thickness and crystallinity, the sensor response spans the full gamut of structural colors, where each color corresponds to a unique experimental state. This direct correlation validates that the chosen thermochromic sensor configuration is well suited to act as a robust feedback mechanism in a thickness regime between $1.3-48.5$~nm (~$2-80$ \ce{MoS2} layers). In the analyzed thickness regime the attenuation and phase shift accumulation in \ce{MoS2} is highly effective and thickness variations in the order of one monolayer can be detected with a simple low-budget spectrometer assuming a spectral resolution of $3$~nm, as discussed in the supporting information (Figure~S2). 

\subsection{Physical properties of \ce{MoS2}}
\begin{figure*}[h!]
\begin{center}
  \includegraphics[width=0.9\linewidth]{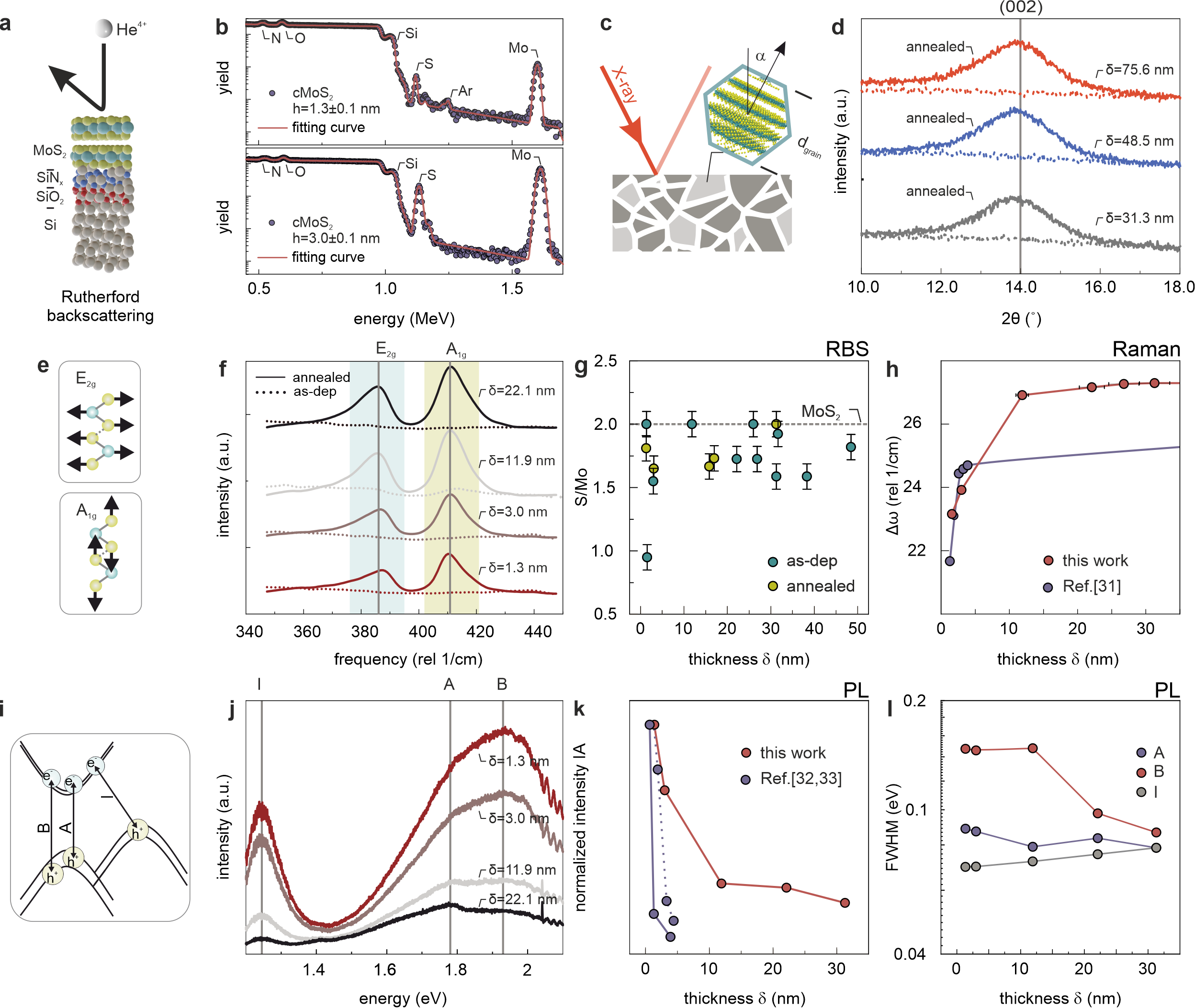}
  \caption{(a)  Schematic diagram of the backscattering process of high energy particles (2~MeV He$^{4+}$) on ultra-thin \ce{MoS2} on passivated silicon substrates. (b) Selected Rutherford backscattering spectra of samples with 1.3~nm (top) and 3~nm (bottom) \ce{MoS2} on top of passivated Si wafers before annealing.  The blue dots are the experimental data, while the orange line indicates the fit from which the parameters were extracted. (c) Schematic illustrating the nanocrystalline nature of the fabricated films. (d) X-ray diffraction pattern of selected thicker \ce{MoS2} films on passivated Si both before (dashed) and after annealing (continuous). The line indicates the position of the c-\ce{MoS2} (002) lattice plane \cite{buck_structure_1986}. (e) Schematic showing the characteristic vibrational modes of TMCs: $E_{2g}$, the in-plane vibration, and $A_{1g}$, the out-of-plane stretching of sulfur atoms. (f)  Raman spectra of sputtered $\ce{MoS2}$ layers on passivated Si both before (dotted) and after annealing (solid line). (g) Mo/S ratio measured by RBS. Blue points correspond to as-deposited samples, while yellow points correspond to annealed samples. (h) Plot of the frequency difference between the two Raman modes shown in (f) for samples with different thicknesses (red). The blue points indicate the equivalent values from the literature \cite{lee_anomalous_2010}. (i) Schematic of the A, B, and I excitonic transitions present in TMCs.
  (j) Photoluminescence spectra of various $\ce{c-MoS2}$ films on passivated Si after annealing. The lines indicate the position of the A, B, and I excitonic transitions. Peak position was determined by peak deconvolution, as shown in Figure S~5. (k) Normalized intensity of the emission of the A exciton against film thickness $\delta$  (red). The blue points indicate the equivalent values from the literature (\cite{splendiani_emerging_2010},\cite{eda_photoluminescence_2011}). (l) Full-width half maximum (FWHM) of the excitonic peaks.}
  \label{fig:MoS2_2}
  \end{center}
\end{figure*}
Having established the relationship between optical response and thickness, 
 we extend the fabrication of \ce{MoS2} to standard passivated silicon wafers (50~nm SiN$_x$| 50~nm SiO$_2$| Si) to better study its physical properties. Additional discussion is provided in the supporting information. We choose Rutherford backscattering spectroscopy (RBS), Raman spectroscopy, photoluminescence spectroscopy (PL), and x-ray diffraction (XRD) to assess the evolution of stoichiometry, crystal structure, characteristic vibrational modes, and excitonic transitions as a function of thickness and post-deposition annealing (Figure~\ref{fig:MoS2_2}). RBS is a reference-free technique that utilizes the element and depth-specific elastic backscattering of high energy particles (Figure\ref{fig:MoS2_2}~a) to extract compositional depth profiles. Due to its sub-nanometer depth and high mass resolution, RBS is ideally suited to probe the stoichiometry, composition, and film thickness in ultra-thin systems. 
\par
Figure~\ref{fig:MoS2_2}~b shows two representative RBS spectra corresponding to two ultra-thin \ce{c-MoS2} films. The stoichiometry and thickness of the \ce{MoS2} layers were obtained by fitting the RBS spectra, and the best fit was achieved for a layer thickness of $1.3\pm 0.1$ and $3.0\pm 0.1$~nm, respectively. The measured average thickness corresponds to systems with 2 and 5 layers and agrees well with those observed in the thermochromic sensors. The spectrum of the thinnest sample ($\delta=1.3$~nm) exhibits a small amount of argon (Ar) impurities, which is a common phenomenon observed in sputter-deposited systems~\cite{GUPTA2022109848}. According to RBS, the average S to Mo ratio is $1.7\pm0.3$ (Figure~\ref{fig:MoS2_2}~g) corresponding to an average sulfur vacancy percentage of $15\%$. We attribute the formation of S-vacancies to the presence of an Ar-plasma during the growth process~\cite{Li2016,GUPTA2022109848}. The presence of such vacancies can lead to the formation of gap states~\cite{Li2016} and can significantly change the physical properties of the material~\cite{alev_nanostructured_2023,donarelli_tunable_2013}. 
\par
We confirm the crystallization of the system upon isothermal annealing using X-ray diffraction (XRD). Figure~\ref{fig:MoS2_2}~d shows XRD-spectra of three selected \ce{MoS2} films, before and after annealing at 750~K. Crystallization of \ce{MoS2} occurs into an in-plane orientation with high nanocrystallinity and average grain size of $d_{\text{grain}}$ of $5.5$~nm (Figure~\ref{fig:MoS2_2}~c,d). The TEM characterization of a film with $\delta$=1.3~nm (Figure~S~5) seems to indicate only partial crystallization. However, this could be due to the low substochiometric Mo/S ratio of the thinnest samples (Figure~\ref{fig:MoS2_2}~g). We refer to the supporting information (Table~4) for details about the crystallite size determination. The formation of such a nanocrystalline phase can lead to additional defect states, such as the accumulation of S-vacancies in the grain boundaries (GBs)~\cite{Sangwan2015}.
\par
To examine the combined effect of chemical and structural disorder observed by RBS and XRD on the vibrational modes of \ce{MoS2}, we resort to Raman spectroscopy. Figure~\ref{fig:MoS2_2}~e illustrates the first-order Raman modes of \ce{c-MoS2}~\cite{zhou_raman_2014,lee_anomalous_2010,conley_bandgap_2013}: the out-of-plane vibration of the sulfur atoms, ($A_{1g}$) and the in-plane vibration of Mo and S atoms ($E_{2g}$). The ($E_{2g}$) mode is an in-plane collective oscillation of Mo and S atoms, and it is therefore susceptible to changes in stoichiometry and in-plane strain. The peaks corresponding to these modes emerge after annealing, as shown in the Raman spectra in Figure~\ref{fig:MoS2_2}~f. Both peaks exhibit asymmetrical broadening, indicating that the presence of small grain sizes induces phonon confinement effects~\cite{PhysRevB.91.195411}. The grain boundaries (GBs) can disrupt spatial translational symmetry, leading to a substantial reduction in phonon lifetime, ultimately resulting in an increased linewidth (Figure~\ref{fig:MoS2_2}~e).
\par
Although a substoichiometric Mo/S ratio has been shown to increase the frequency difference between the two Raman modes~\cite{donarelli_tunable_2013,parkin_raman_2016,kim_changes_2018,alev_nanostructured_2023}, we still observe the characteristic shift of the first-order modes as a function of thickness $\delta$ (Figure~\ref{fig:MoS2_2}~h). The measured frequency difference $\Delta \omega$ gets smaller in the few-layer regime. This shift indicates that, despite the high density of defects, the changes in the phononic band structure with the reduction in the number of layers is comparable to that observed for \ce{MoS2} deposited by other techniques~\cite{lee_anomalous_2010,tao_growth_2015,mazaheri_mos2--paper_2020}.
\par
To extend the analysis towards the optoelectronic properties, which are also known to be thickness-dependent~\cite{splendiani_emerging_2010}, we performed steady-state photoluminescence spectroscopy on \ce{c-MoS2}. Inspection of the obtained PL spectra (Fig.~\ref{fig:MoS2_2}~j) reveals three excitonic emission peaks, which can be attributed to indirect-gap luminescence (peak I $1.24$~eV), and direct-gap hot luminescence (peak A $1.78$~eV, peak B $1.93$~eV), as illustrated in Figure~\ref{fig:MoS2_2}~i~\cite{mak_atomically_2010,leonhardt_use_2020}. In case the \ce{MoS2} thickness exceeding one monolayer, the direct-gap transition occurs due to hot carriers that transiently occupy spin-orbit split valence bands near the K-points~\cite{mak_atomically_2010}. It is to be noted that in addition to direct-gap hot luminescence demonstrated here, contributions of directly thermalized carriers from the fundamental band gap cannot be fully ruled out and have to be considered to contribute to the radiative excitonic emission.
\par
The linewidth of the A-, and B- exciton are significantly broadened, so that the two peaks cannot be distinguished without peak deconvolution (see the supporting information, Figure~S~6 and Table~3). As the linewidth of the PL signal is inversely correlated with the excitonic lifetime $\tau$~\cite{Ajayi_2017, selig_excitonic_2016,kira_many-body_2006}, it is a clear indicator of non-radiative recombination due to S-vacancies or structural disorder~\cite{mccreary_-_2018,tongay_defects_2013,nan_strong_2014}. The short excitonic lifetime makes sputtered \ce{MoS2} films strong candidates for valleytronics~\cite{mccreary_-_2018}. 
\par
The high measured $B/A$ peak ratio (Fig.~\ref{fig:MoS2_2}~k) and the increasing linewidth of the B-exciton with decreasing film thickness (Fig.~\ref{fig:MoS2_2}~l)~\cite{mccreary_-_2018} further indicate increase defects in thinner films~\cite{mccreary_-_2018,kaupmees_photoluminescence_2019,wang_monitoring_2022}. The intensity of direct-gap excitons $A$ and $B$ scales inversely with the film thickness $\delta$ (Figure~\ref{fig:MoS2_2}~k), reminiscent of exfoliated systems~\cite{splendiani_emerging_2010}. This can be attributed to the direct-to-indirect bandgap transition and/or to increased defect density~\cite{tongay_defects_2013,nan_strong_2014}. The persistence of the emissivity of these direct-gap excitons in films with thicknesses of tens of nanometers contrasts with \ce{MoS2} fabricated through other methods. The stoichiometric and lattice defects could provide recombination channels allowing for the characteristic scaling of the emission of the $A$ and $B$ excitons.
\par
It is remarkable that, despite this nano crystallinity and the sub-stoichiometry in our films, we still observe the same trends in the scaling of the electronic and phononic bandstructure with thickness as in \ce{MoS2} obtained by other methods such as CVD or exfoliation. The notion that the physical properties prevail over their thickness dependence despite the high-defect concentration, highlights the robustness of sputtered \ce{MoS2} films
\subsection{Amorphous-Crystalline Transition}
\begin{figure*}[h!]
\begin{center}
  \includegraphics[width=0.9\linewidth]{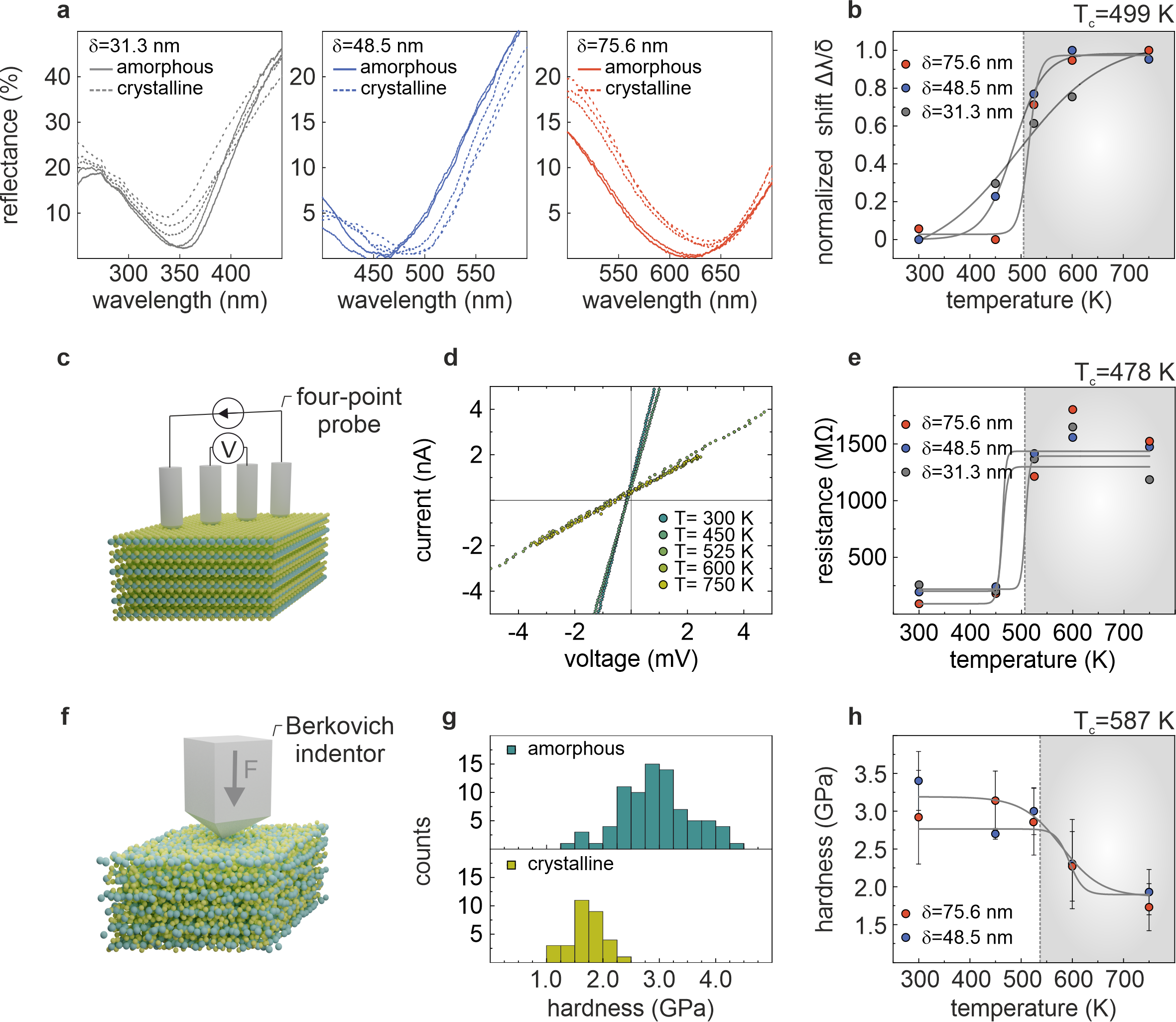}
  \caption{(a) Near-normal incidence reflectance spectra of three selected $\ce{MoS2}$ systems with characteristic dip in reflection. The solid lines show the spectra of the amorphous state and the dotted lines show the reflectance spectra of the samples annealed at 450~K, 525~K, 600~K, and 750~K. The spacing of the dots increases with the annealing temperature. (b) Change of the minimal reflectance extracted from (a) given by $\Delta \lambda/\delta$ at different annealing temperatures. The grey shaded area indicates the region above the calculated crystallization temperature $T_c$. All curves exhibit a step-wise transition around $T_c$ as fitted using a modified hyperbolic function for each $\ce{MoS2}$ thickness.(c) Schematic of the four-point probe used to measure the sheet resistance. (d) Current-voltage curves as measured by a four-point probe for the different crystallization states of the 48.5~nm film. (e) Evolution of the sheet resistance of selected $\ce{MoS2}$ systems across the amorphous-crystalline phase transition. The grey area indicates the average $T_c$, as fitted using a modified hyperbolic function for each $\ce{MoS2}$ thickness. (f) Schematic of the nanoindentation setup used to measure hardness. (g) Histograms showing the distribution of the measured hardnesses for the $\delta=48.5$~nm film. The blue histogram at the top is for the as-deposited state, while the red histogram below was measured on the same sample annealed to 750~K. (h) Hardness of the samples annealed at different temperatures. The grey area indicates the average $T_c$, as fitted using a modified hyperbolic function for each \ce{MoS2} thickness.}
  \label{fig:MoS2_3}
\end{center}
\end{figure*}
So far, we limited the analysis of the functional properties to a mere comparison of one amorphous and one crystalline state. An interesting question is can one use the direct optical feedback to track the change of functional properties across the amorphous-crystalline transition. Such an approach can allow for valuable insights, including the determination of the crystallization temperature $T_c$ and identification of metastable intermediates~\cite{kolobov_structural_2020,Kitchaev2016}.
\par
Figure~\ref{fig:MoS2_3}~a shows near-normal incidence reflectance spectra of three \ce{MoS2} thin films with  $\delta=$~31.1~nm, $\delta=$~48.5~nm and $\delta=$~75.6~nm sputtered on SiN$_x$/SiO$_2$/Si substrates and annealed at different temperatures. All spectra exhibit a characteristic minimum in reflectance which shifts upon annealing due to the change in the optical properties of \ce{MoS2}. The inversion of the wavelength shift in Figure~\ref{fig:MoS2_3}~a is due to the refractive index of the passivated silicon wafers being higher than that of \ce{MoS2} for the $\delta=$~31.1~nm film minimum and being lower for the other two systems. Figure~\ref{fig:MoS2_3}~b illustrates the absolute value of the normalized relationship between $\lambda_{min}$ and the thickness of the film. We observe a typical step-like function indicative of a first-order transition, with the average transition temperature $Tc=499\pm99$~K. It should be noted that 750~K is the temperature at which the samples in Figures~\ref{fig:MoS2_1} and \ref{fig:MoS2_2} were annealed and that the XRD results shown in Figure~\ref{fig:MoS2_2}~b correspond to the same batch of samples. Therefore, we attribute this shift in the minimal reflectance to crystallization and show that, for thicker films, the change in the optical properties of \ce{MoS2} upon crystallization can be sensed without a metallic backreflector, opening the door to characterize the electrical properties of \ce{MoS2}.
\par
We evaluated the electrical properties of our films using a four-point linear probe, as illustrated in Figure~\ref{fig:MoS2_3}~c. In Figure~\ref{fig:MoS2_3}d, the ohmic current-voltage characteristic of the $\delta=$~48.5~nm film is depicted, illustrating an increase in sheet resistance with varying annealing temperatures, with a noticeable increase marked by the dashed line upon crystallization. The increase in sheet resistance was similarly observed for the other two film thicknesses, as depicted in Figure~\ref{fig:MoS2_3}~e. The average fitted transition temperature ($Tc=478\pm93$~K) is in good agreement with the transition temperature found through the characterization of the optical properties (Table~S5). An increase in sheet resistance of almost one order of magnitude upon annealing could be measured for film thicknesses down to $11.9$~nm, as shown in Figure~S7. We attribute this stark increase in sheet resistance upon crystalization to changes in the coordination of molybdenum atoms. The metallic homopolar Mo-Mo bonds in the amorphous state are replaced by less conductive Mo-S covalent bonds upon crystalization~\cite{krbal_amorphous--crystal_2021,krbal_anomalous_2023}. 
\par
To characterize the mechanical properties of our films, we performed nanoindentation experiments, as illustrated by the schematic in Figure~\ref{fig:MoS2_3}~f. We observe a decrease in the hardness of the films after annealing, as shown in Figure~\ref{fig:MoS2_3}~g and h. This behaviour can also be explained through the amorphous-crystalline phase transition in the film. Intramolecular bonds dominate the atomic structure of \ce{a-MoS2} and typically lead to brittle behaviour and high hardness~\cite{cai_imperfections_2016}. On the contrary, the stacked structure of \ce{c-MoS2} is characterized by strong covalent in-plane forces and weak interlayer van der Waals interactions. These interplane van der Waals interactions provide low resistance to applied forces, enabling the sliding or shearing of layers at substantially lower stress than for intramolecular bonds~\cite{serles_high_2022}. Interestingly, interlayer sliding also enables the superior lubrication of \ce{MoS2}~\cite{rapoport_friction_2008,serles_high_2022}. Therefore, during hardness measurements, the weak out-of-plane bonding in these stacked sheet structures reduces the hardness of the material by lowering the stress required for deformation and reaccommodation within the structure.
\par
Furthermore, we observed a large peak broadening in the XRD measurements that indicates substantial out-of-plane rotations between the crystalline \ce{MoS2} grains and the substrate (schematically illustrated in Figure~\ref{fig:MoS2_2}~j). These grain rotations make interlayer sliding even more accessible as a deformation mechanism under indentation loading and would further support the mechanical availability of interplane sliding. Thus, in the presence of a nanocrystalline phase, we expect a clear decrease in hardness as compared to the amorphous phase. The calculated crystallization temperature ($T_c=587\pm106$) is higher than the one found through electrical and optical characterization but falls within the error range. 

\subsection{Few-layer \ce{MoS2} multilayer composites}
\begin{figure*}[h!]
\begin{center}
 \includegraphics[width=0.9\linewidth]{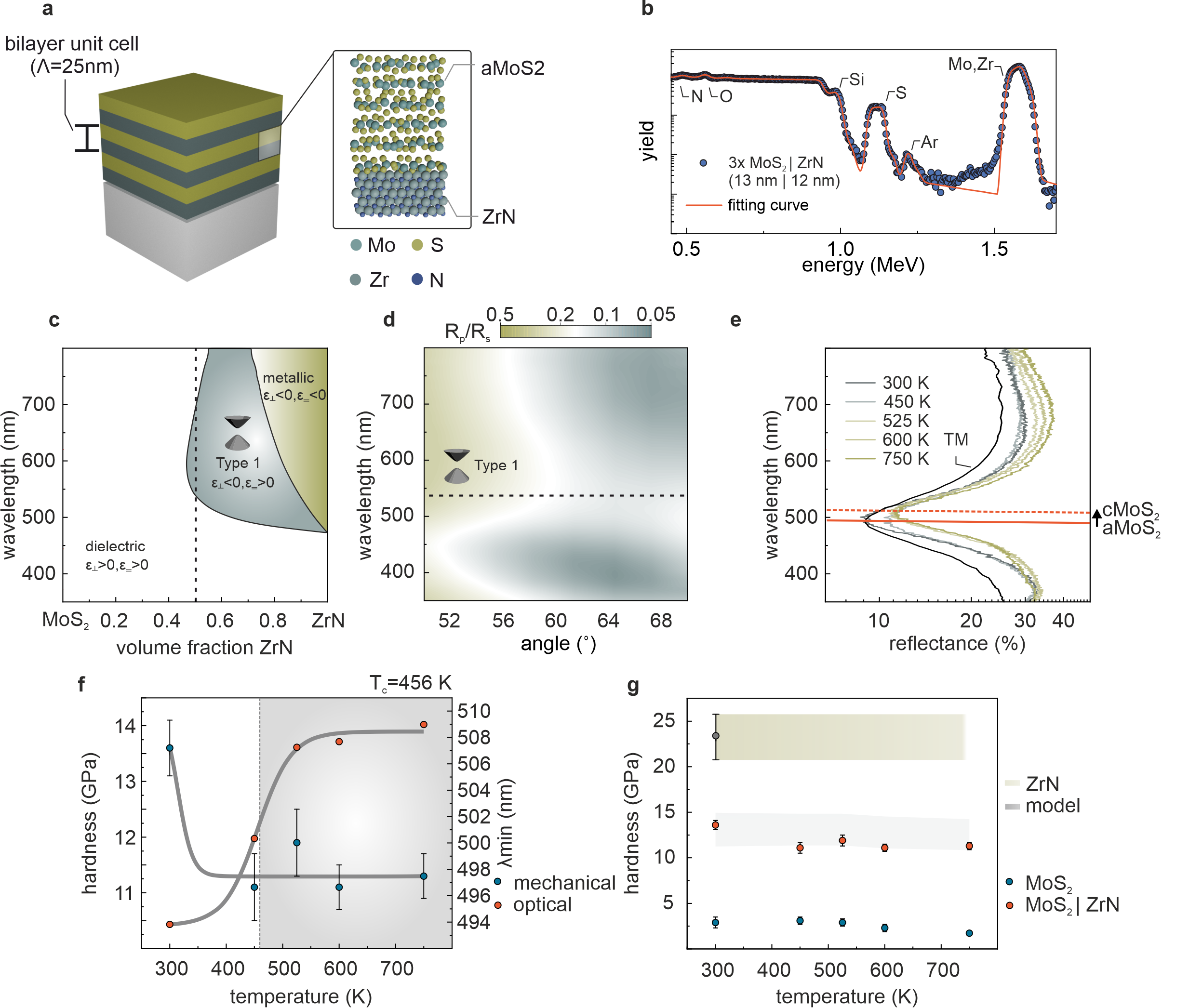}
\caption{(a) Schematic of the deposited \ce{ZrN}-\ce{MoS2} multilayer. (b) Rutherford backscattering spectra of the \ce{ZrN}-\ce{MoS2} multilayer. The blue dots correspond to the experimental data, while the orange line indicates the simulated spectrum. (c) Optical phase diagram for \ce{MoS2}-\ce{ZrN} HMMs. The colour code depicts the different zones: effective dielectric ($\epsilon_\perp> 0, \epsilon_\parallel > 0$ ), effective metal ($\epsilon_\perp< 0, \epsilon_\parallel < 0$), and Type I HMM ($\epsilon_\perp < 0$, $\epsilon_\parallel > 0$). The dashed vertical lines illustrate the measured composition. (d) Color mapping of the measured ratio in reflectance between the p- and s-polarization. The dashed line indicates the wavelength at which the transition between a dielectric and an HMM is expected. (e) Near-normal incidence reflectance spectra of the HMM annealed at different temperatures. The dashed dark line indicates the calculated reflectance using the TM method. The horizontal red lines indicate the wavelength of the HMM with amorphous (solid) and crystalline (dashed) \ce{MoS2}, respectively. (f) Hardness of the multilayer metamaterial (red) and \ce{MoS2} thin film (blue) annealed at different temperatures. The semi-transparent yellow area shows the measured hardness of \ce{ZrN}. The semi-transparent grey area indicates the expected results calculated using the rule of mixtures. (g) Change in the hardness (blue) and wavelength of the reflected minimum (red) of the multilayer metamaterial with the temperature of annealing. The grey area indicates the temperatures after which \ce{c-MoS2} is expected, as fitted from the optical data with a modified hyperbolic function.}
  \label{fig:MoS2_4}
  \end{center}
\end{figure*}
The ability to rapidly track the thickness and phase of sputter-deposited \ce{MoS2} opens the possibility of creating more complex stacked photonic systems combining \ce{MoS2} with non-van der Waals materials. We illustrate this by fabricating a \ce{MoS2}/\ce{ZrN} hyperbolic metamaterial~\cite{ocana,cortes_quantum_2014}. These architected materials enable the engineering of the local density of optical states (LDOS), enhancement of spontaneous emission, and confinement of light to small modal volumes~\cite{cortes_quantum_2014}. Furthermore, hyperbolic metamaterials are considered as a promising platform to create plexcitons, strongly coupled plasmon-excitonic quasi-particles~\cite{zheng_quantum_2023}.
\par
Figure~\ref{fig:MoS2_4}~a illustrates the metamaterial design consisting of alternating layers of \ce{ZrN} and \ce{MoS2}. The bilayer unit cell $\Lambda$ and the volume fraction $\phi$ were chosen to be $25$~nm and $0.5$, respectively. The composition and thickness profile of our fabricated \ce{ZrN}/\ce{MoS2} HMM was measured by RBS and is shown in Figure~\ref{fig:MoS2_4}~b. Due to the small energy separation between \ce{Mo} and \ce{Zr}, the alternation of the layers is only resolvable within the simulation of the spectrum.
\par
According to effective medium theory, the chosen metamaterial design exhibits an optical topological transition (OTT) in the visible spectrum (Figure~\ref{fig:MoS2_4}~c)\cite{ocana,krishnamoorthy_topological_2012,poddubny_hyperbolic_2013}. Due to the emergence of new optical states, the isofrequency surface of the HMM changes its shape from a sphere (dielectric phase) to a hyperbole (Type 1 phase). We experimentally observe this transition between the two topologies by measuring the $R_{p}/R_{s}$ ratio (Figure~\ref{fig:MoS2_4}~d), since the propagation of hyperbolic modes is limited to p-polarized light~\cite{hoffman_negative_2007,ferrari_hyperbolic_2015}. The transition from the dielectric phase to the Type I HMM phase is due to a change of sign in the out-of-plane permittivity  $\epsilon_\perp$ and results in a discontinuity in the pseudo-Brewster angle (\cite{hoffman_negative_2007,cho_experimental_2021}). Our measurements replicate this discontinuity at $\lambda=$525~nm, as shown in Figure~\ref{fig:MoS2_4}~d.
\par
When annealed across the amorphous-crystalline transition, the optical properties of the HMM change. Figure~\ref{fig:MoS2_4}~e compares the measured reflectance at normal incidence of HMMs annealed at different temperatures. All HMMs exhibit a Lorentz-like dip in reflectance which is caused by coupling to propagating modes with high-k wavevector. The reflectance minimum redshifts upon annealing(Figure~\ref{fig:MoS2_4}~f), and the shift reproduces the characteristic profile of the phase transition as observed in Figure~\ref{fig:MoS2_3}.
\par
Understanding the composite's architecture also enables the design of the multilayer's mechanical properties. We observe that the hardness of the \ce{MoS2}-\ce{ZrN} multilayer decreases upon annealing, following the same trend as the hardness of \ce{MoS2}. The mechanical properties of \ce{MoS2} can be deconvoluted using the rule of mixtures, as illustrated in Figure~\ref{fig:MoS2_4}~f. The hardness value of \ce{ZrN} is taken as the mean hardness measured in the as-deposited samples, while the value for $H_{MoS_2}$ comes from the hardness measurements shown in Figure~\ref{fig:MoS2_3}~h. At 50\% volume fraction \ce{MoS2}, the calculated hardness for the as-deposited samples is in good agreement with the measured hardness. Thus the composite mechanical response of the HMM does not indicate strong size effect contributions arising from the ultra-thin layers. Additionally, mechanical degradation in \ce{ZrN} after annealing could contribute to a reduction in multilayer hardness that is not reflected in the calculated values. Another possible contributing factor could include the relative fraction of probed materials~\cite{choi_nanoindentation_2009}. As \ce{c-MoS2} is softer than \ce{a-MoS2}, it will have a larger plastic zone and therefore a larger contributing volume to the probed material under indentation conditions. This would increase the relative volume fraction of \ce{MoS2} and thus reduce the predicted hardness.
\par 
The results demonstrate that the rule of mixtures can be used to design the hardness of layered metamaterials, in a similar fashion to how the effective medium approximation can be used to design the optical phase diagram of the multilayer, opening the door to the design of hyperbolic metamaterials with tailored mechanical properties. With an order of magnitude difference in hardness between the constituents, \ce{MoS2}/\ce{ZrN} multilayers provide a large space to tailor the mechanical response to the distinct hardnesses. Remarkably, the applicability of the rule of mixtures allows for rapid prototyping and design in comparison to first-principle \cite{calzolari_hyperbolic_2021}.
\par
The multilayer can also be used to determine the hardness of \ce{MoS2} $\delta=$~12.5~nm ($~20$ monolayers) in a deconvoluted manner. In addition to enabling sensing the amorphous-crystalline phase transition and displaying hyperbolic optical modes in the visible, the \ce{ZrN}/\ce{MoS2} multilayer shows the characteristic \ce{MoS2} photoluminesce response, as shown in Figure~S9 in the supporting information. These stacked designs pave the way for the realisation of plexitonic devices with tailored mechanical properties, enabling the study of the effect of strain in the exciton-plasmonic quasiparticles, among others.

\section{Conclusion}

The present work highlights the potential of smart sensing to determine the phase, thickness, and functional properties of \ce{MoS2} fabricated by bottom-up magnetron sputtering. The sensors can detect \ce{MoS2} thickness variations down to one monolayer and display a full gamut of colors obtained by exploiting the freedom in substrate choice and control over thickness and phase. The coloration can be used as a fast post-deposition feedback mechanism, opening the door to further controlling the thickness and crystallinity through \ce{MoS2} fabrication.
\par
We show that the effects of the phase change upon crystallization are crucial for understanding the phononic and electronic band structures of our films. For example, we observe that even in bulk-like samples, the direct-gap hot luminescence typically associated with monolayer \ce{MoS2} dominates the PL spectra. We link the unexpected enhancement in quantum yield to nanocrystallinity and related finite size effects on the excitonic emission. Despite the defects induced by our fabrication technique, the electronic and phononic band structures of our films shift with thickness in a way similar to \ce{MoS2} produced by exfoliation or CVD. 
\par
Leveraging the changes in optical properties through crystallization, we explore the accompanying changes in electrical and mechanical properties of \ce{MoS2} through the phase transition. We further demonstrate that the valuable insights from this optical sensing can be garnered to fabricate an \ce{MoS2}-based HMM that allowed us to detect changes in the hardness of few-layer \ce{MoS2} with a simple low-budget spectrometer. The hardness of these multilayers can be approximated with the rule of mixtures, opening the door to the rapid design of hyperbolic metamaterials with tailored mechanical properties. Additionally, the demonstrated presence of excitonic features in the \ce{MoS2}-based multilayers paves the way for the study of interesting physical phenomena such as the effect of strain plexitonic modes.
\par
Our findings provide insights into the functional properties of amorphous and crystalline \ce{MoS2} across different thicknesses and show a viable route towards versatile wafer-scale fabrication of \ce{MoS2} thin films and \ce{MoS2}-based architectures, thus contributing to the development of \ce{MoS2}-based devices. 
\section{Experimental Section}
\threesubsection{Synthesis and annealing}
The materials in this work were deposited using magnetron sputtering (\emph{PVD Products Inc}). Molybdenum disulfide (99\% \emph{MaTecK GmbH}) was sputtered by radio frequency (RF) sputtering, while zirconium nitride was deposited by reactive sputtering using a zirconium target (99.5\% \emph{MaTecK GmbH}) and constant nitrogen flow (3 sscm). The samples were annealed for 2~h in a vacuum at 1 $\times$ 10$^{-9}$ mbar in a \emph{Createc} rapid-thermal annealing (RTA) setup. The heating rate was kept at 5~K/min and no active cooling was used.
\par
\threesubsection{Optical characterization and calculations}
The refractive index of \ce{MoS2} and \ce{ZrN} was measured by Variable Angle Spectroscopic Ellipsometry (VASE) with a \emph{J.A. Woollam} M-2000 system. A comparison between the experimental n and k values of magnetron sputtered \ce{MoS2} and \ce{ZrOxNy} with data from the corresponding literature is shown in Figure~S3, supporting information. M-2000 was also used to measure the angular and polarization-dependent reflectance spectra. The reflectance spectra at near-normal incidence were measured using a fiber-coupled reflectometer (OceanOptics). Transfer Matrix (TM) calculations were performed on a \emph{Wolfram Mathematica}. Photoluminescence measurements were performed on a Horiba microscope (LabRAM HR Evolution UV-VIS-NIR). A 532nm Nd: Yag laser was used at a nominal power of 30~W and focused onto the samples with a Nikon PlanFluor 10x objective. The incident power was controlled through the filter wheel setting set at 10\% after observing no shift in the optical response below that threshold. The integration time was kept constant to 45~s.
\par
\threesubsection{Structural characterization}
The X-ray diffraction (XRD) measurements were performed using an (X’Pert MRD, Panalytical, Netherlands) equipped with a 0.27 $^\circ$ parallel plate collimator. $Cu K\alpha1$ ($\lambda = 1.540598$$\textup{~\AA}$ radiation generated at 40 kV/45 mA was used. RBS measurements were performed with a 2 MeV 4He beam and a Si PIN diode detector under $168^\circ$ scattering angle. The results were fitted using RUMP. The Raman measurements were performed on a Witec Microscope CRM 200 with a 532~nm excitation. All measurements were performed in backscattering mode with an edge filter long pass 532~nm filter. A 20x objective with a numerical aperture of 0.4 was used. We refer to the supporting information for the details on the peak deconvolution (Table~S3). The integration time was kept constant at 30~s. 
\par
\threesubsection{Electrical and mechanical characterization}
The sheet resistance was measured using a linear four-point probe head with tungsten electrodes and a Keysight B2962A power supply. The current-voltage curves were measured in the Ohmic region, varying from nA to mA. An iNano nanoindenter (Nanomechanics, Inc.) with an InForce50 actuator under a quasistatic strain rate of 0.1~s$^{-1}$ using a diamond Berkovich indenter tip (Synton-MDP) was used for the indentation experiments.  All \ce{MoS2} thin film samples were indented to a depth of 20~nm in a 10x10 array. The load-displacement curves were inspected for artefacts and punch-through of the films, as indicated by a discrete shift in loading slope to identify successful indentation experiments. To appropriately account for the bluntness of our indentation tip at such small indentation depths, a spherical approximation with a 50~nm radius of curvature was made for the contact area. The \ce{MoS2}/\ce{ZrN} multilayers were indented to a load of 1~mN in a 10x10 array to allow the penetration of multiple layers for all samples. The \ce{ZrN} multilayers were indented in a 10x10 array to a load of 1~mN in four regions of the sample, leading to an average indentation depth of $47$~nm. All hardness measurements are reported at the maximum indentation depth.

\textbf{Supporting Information} \par 
Supporting Information is available.\par
\textbf{Acknowledgements} \par 
The authors acknowledge the infrastructure and support of FIRST and ScopeM. The authors acknowledge the valuable help from Yuting Chen, Joan Sendra, Carmen Launer, Hekun Kuang, and Amit Korde; and the discussions with Jelena Wohlwend, Reindhard Kaindl, and Fabio Krogh. J.L.O.P thanks for the financial support from the ETH research grant (ETH-47 18-1). 
\bibliography{bibliography_mos2}
\end{document}